\documentclass[%
reprint,
superscriptaddress,
 amsmath,amssymb,
 aps
]{revtex4-1}

\usepackage{amsmath}
\usepackage{algorithm}
\usepackage[noend]{algpseudocode}

\makeatletter
\def\BState{\State\hskip-\ALG@thistlm}
\makeatother

\usepackage{graphicx}
\usepackage{dcolumn}
\usepackage{bm}
\usepackage{booktabs}
\usepackage{amsmath}
\usepackage[flushleft]{threeparttable} 

\usepackage{dsfont}
\usepackage{color,psfrag}
\usepackage{mathtools,amscd}

\usepackage{soul}
\usepackage{xcolor} 
\definecolor{purple}{rgb}{0.5,0.0,0.5}
\definecolor{mypink1}{RGB}{219, 48, 122}
\definecolor{mypink2}{cmyk}{0, 0.7808, 0.4429, 0.1412}
\definecolor{mygray}{gray}{0.6}
\definecolor{mygreen}{HTML}{0CA172}
\definecolor{myblue}{HTML}{20517D}

\newcommand\here[1]{\fcolorbox{red}{red}{\rule{0pt}{6pt}\rule{6pt}{0pt}}\quad}
\newcommand\missing[1]{\fcolorbox{blue}{blue}{\rule{0pt}{6pt}\rule{6pt}{0pt}}\quad}

\usepackage{ifpdf}
    \ifpdf
\usepackage{tikz}
\usetikzlibrary{arrows,chains,matrix,positioning,scopes, fit, calc}
\usetikzlibrary{cd}
\usepackage{chemfig}
\fi
\usepackage{hyperref}

\hypersetup{
	colorlinks,
	linkcolor=blue,
	citecolor=blue, 
	filecolor=black,
	urlcolor=blue}

\begin{document}

\preprint{APS/123-QED}


\title{Towards Accelerated SCF Workflows with Equivariant Density-Matrix Learning and Analytic Refinement}

\author{Zuriel Y. Yescas-Ramos}
\affiliation{Instituto de F\'isica, Universidad Nacional Aut\'onoma de M\'exico, Cd. de M\'exico C.P. 04510, Mexico}

\author{Andr\'es \'{A}lvarez-Garc\'ia}
\affiliation{Instituto de F\'isica, Universidad Nacional Aut\'onoma de M\'exico, Cd. de M\'exico C.P. 04510, Mexico}

\author{Huziel E. Sauceda}
\email{huziel.sauceda@fisica.unam.mx}
\affiliation{Instituto de F\'isica, Universidad Nacional Aut\'onoma de M\'exico, Cd. de M\'exico C.P. 04510, Mexico}

\date{\today}

\begin{abstract}
We present \textsc{dm-PhiSNet}, a physically constrained \textsc{PhiSNet}-based equivariant model that predicts one-electron reduced density matrices (1-RDMs) directly from molecular geometries in an atomic-orbital (AO) basis for accelerated self-consistent field (SCF) workflows. Training follows a two-stage schedule with progressively introduced physically motivated objectives, and the resulting predictions are refined by a lightweight analytic block. This block enforces electron-number conservation, drives the 1-RDM toward generalized idempotency in the AO metric, and regularizes the occupation spectrum of the L\"owdin-orthogonalized density. Across six closed-shell systems---H$_2$O, CH$_4$, NH$_3$, HF, ethanol, and NO$_3^-$---the refined 1-RDMs provide SCF initial guesses that substantially reduce iteration steps by 49--81\% relative to standard initializations. Beyond SCF acceleration, the learned 1-RDMs yield accurate one-shot total energies and Hellmann--Feynman atomic forces without force supervision, indicating that the model captures chemically meaningful electronic structure. These results demonstrate that combining equivariant learning with analytic constraint enforcement provides a simple, general route to solver-ready density-matrix initializations and accelerated SCF workflows.
\end{abstract}

\maketitle

\vspace{0.35em}

\section{Introduction}
\label{introduction}

Machine learning (ML) is increasingly used in electronic-structure theory to emulate electronic quantities---from electron densities to operator-valued objects such as Hamiltonians, Fock matrices, and one-electron density matrices (1-RDMs)---to reduce or bypass the cost of self-consistent field (SCF) procedures.~\cite{Westermayr2021PerspectiveML,Chandrasekaran2019LearningDFT}
Given reference quantum-chemistry data, such models learn mappings from molecular structure to electronic observables, enabling either one-shot predictions or improved initial guesses.\cite{Westermayr2021PerspectiveML}

For closed-shell restricted Kohn--Sham (RKS) theory in a non-orthogonal atomic-orbital (AO) basis, an initial 1-RDM ($\mathbf{P}_0$) is \emph{physically admissible} for downstream electronic-structure calculations if it (i) conserves the electron number, (ii) satisfies generalized idempotency in the AO metric, and (iii) yields a physically sensible occupation spectrum when expressed in an orthogonalized representation.~\cite{Westermayr2021PerspectiveML,Hu2010ADIIS,Cances1999SCF}
In practice, the complexity faced by SCF solvers in converging an electronic-structure calculation is mainly driven by the admissibility of the initial guess rather than by small entrywise errors in $\mathbf{P}_0$.~\cite{Hu2010ADIIS,Mazziotti2007,Lowdin1950NonOrthogonality}
Not fulfilling this condition can induce oscillations, slow convergence, or divergence of Roothaan--Hall iterations, motivating stabilizers such as damping and DIIS-type schemes.\cite{Hu2010ADIIS,Cances1999SCF}

ML electronic-structure modeling has advanced rapidly in recent years. Early work focused on learning electron densities, from AO-based wavefunction representations such as \textsc{SchNOrb}\cite{Schutt2019SchNOrb} to transferable density-reconstruction models in the gas phase and condensed phase,\cite{Grisafi2019ElectronDensity,Lewis2021CondensedPhaseDensity} enabling both property prediction from learned densities\cite{Grisafi2023PropertiesFromDensity} and density-based DFT emulation.\cite{Ryczko2018DeepDFTDensity,DelRio2023DeepDFTEmulator,2012_LearnFunctionals_Burke,2020_MLDFT_Bogojeski}
Related approaches have explored alternative density parameterizations to improve accuracy and efficiency.\cite{2022_MLinDFT,Focassio2023JacobiLegendre}
In parallel, the scope of ML surrogates has expanded to operator-level targets, including symmetry-aware Hamiltonian learning,\cite{Li2022DeepH,Gong2023DeepH_E3,Zhang2022EquivariantHamiltonianMapping,Zhong2023EquivariantGNNHamiltonians} explicit Fock-matrix construction,\cite{Liu2025MachineLearnedFockMatrix} and differentiable ML--QC pipelines that couple learned components to gradient-based solvers.\cite{Suman2025DifferentiableMLQC}
Overall, the field is moving beyond scalar-property regression toward learning electronic objects under symmetry and physics constraints.\cite{Westermayr2021PerspectiveML}

Within this landscape, \textsc{PhiSNet} introduced an $\mathrm{SE}(3)$-equivariant AO framework for predicting wavefunctions and densities,\cite{Unke2021PhiSNet} and a growing body of work has begun to target the 1-RDM as a compact representation of electronic structure.\cite{Shao2023OneElectronRDM,Hazra2024Predicting1RDM,Rana2024Learning1RDM,Zhang2024DeepDMRefinement}
Many existing 1-RDM predictors, however, are based on dense or linear architectures that do not build rotational equivariance into the model, which can reduce data efficiency and physical interpretability.\cite{Westermayr2021PerspectiveML}
Here, we devise the model \textsc{dm-PhiSNet} to learn one-electron density matrices $\mathbf{P}$ in an atomic-orbital (AO) basis representation. The model comprises two macro-blocks: an $\mathrm{SE}(3)$-equivariant \textsc{PhiSNet} backbone that predicts a raw AO 1-RDM from the molecular geometry, and a lightweight \textsc{AnalyticalBlock} that enforces key admissibility conditions, including electron-number conservation and generalized idempotency in the AO metric.
As a result, the refined 1-RDMs act as robust SCF initial guesses across chemically diverse closed-shell molecules, reducing iteration counts by up to $\sim 80\%$ while also yielding accurate single-shot Hellmann--Feynman atomic forces without force supervision.

\begin{figure*}[t]
\centering
\includegraphics[width=\textwidth]{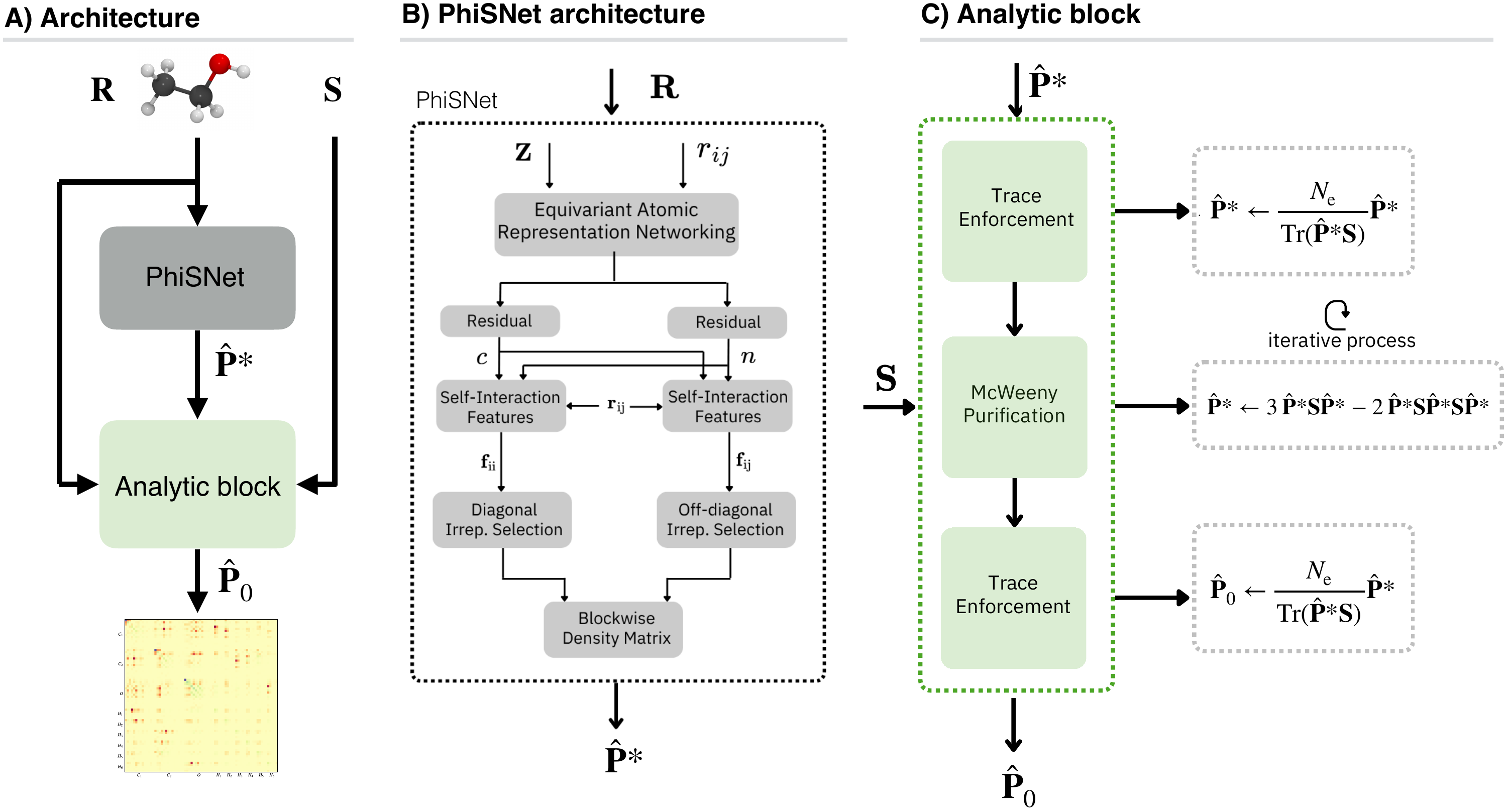}
\caption{\textsc{dm-PhiSNet} model.
(A) The \textsc{PhiSNet}-based backbone (gray box) builds blockwise AO 1-RDMs from molecular geometries $\mathbf{R}$ and atomic numbers $\mathbf{Z}$, which are then refined by the analytic block (green box) to give the 1-RDM predictor.
(B) Representation of the \textsc{PhiSNet} architecture.~\cite{Unke2021PhiSNet}
(C) The \textsc{AnalyticBlock} is constructed by three sub-blocks to enforce physical consistency using the overlap matrix $\mathbf{S}$: (i) trace rescaling, (ii) McWeeny purification cycle, and (iii) a final trace rescaling.}
\label{fig:flowchart}
\end{figure*}

\begin{figure*}[t]
\centering
\includegraphics[width=\textwidth]{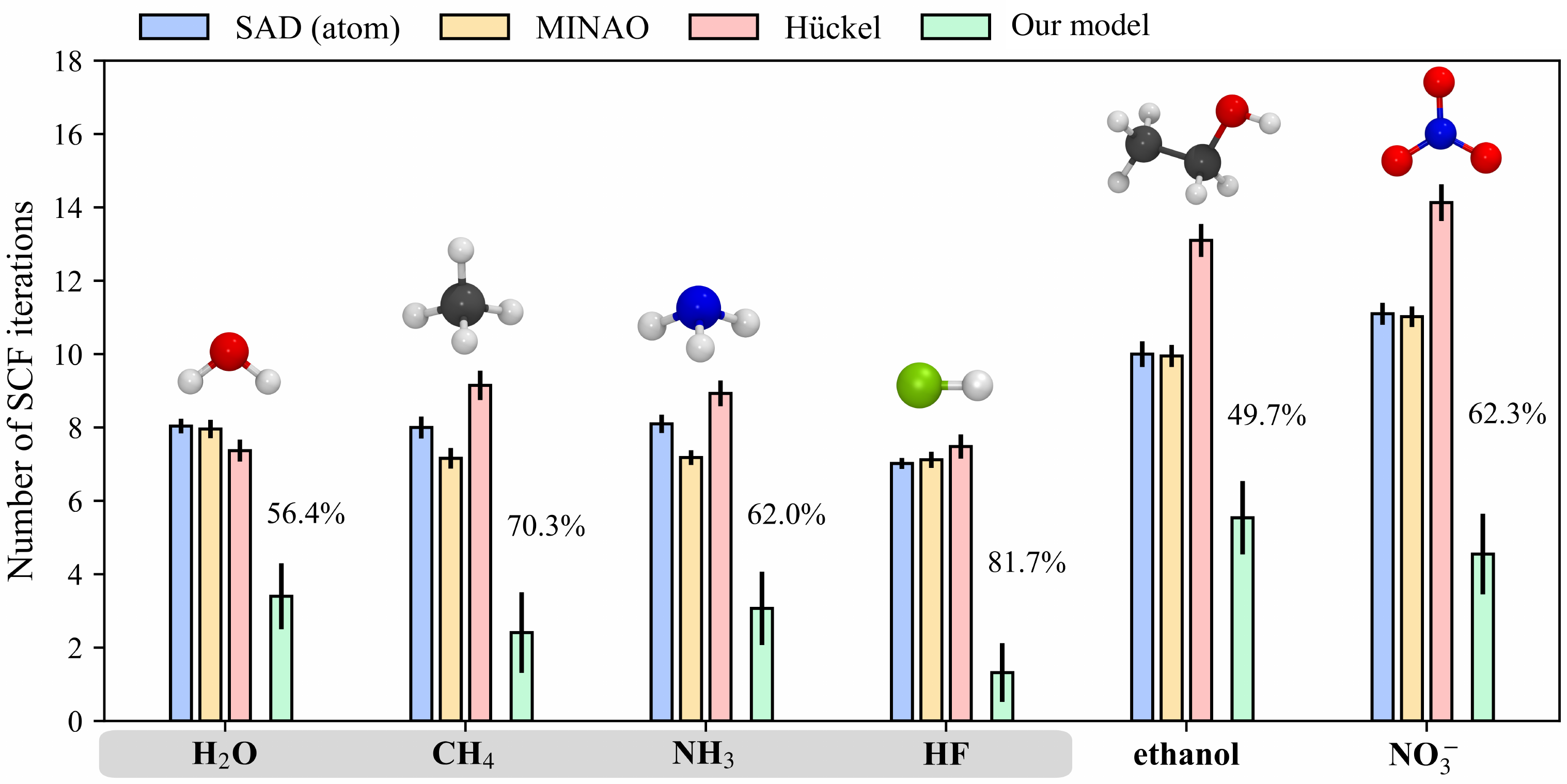}
\caption{Mean numbers of SCF iterations over 500 test configurations for all benchmarked systems.
Results compare the performance of the 1-RDMs predicted by our model against the conventional SAD (\texttt{atom}), MINAO, and H\"uckel initial guesses. Magnitudes of the standard deviations (STD) are represented by vertical black lines on the frequency bars. Reductions with respect to the three averaged standard initial guesses are also shown.}
\label{fig:scf_reductions}
\end{figure*}

\section{Methods}
\label{sec:methods}

The main goal of this work is to provide a starting-guess one-electron density matrix ($\hat{\mathbf{P}}_0$) that is not only close to ground-state 1-RDMs ($\mathbf{P}_{\text{GS}}$) entrywise, but also \emph{physically admissible} for downstream electronic-structure workflows. In practice, SCF convergence is governed less by small Frobenius-norm errors than by whether a candidate $\mathbf{P}_0$ satisfies the defining constraints of a closed-shell 1-RDM in a non-orthogonal AO metric: correct electron number, idempotency, and a sensible occupation spectrum. When these conditions are not fulfilled, even numerically accurate predictions can trigger oscillatory or unstable Roothaan--Hall iterations.
Therefore, a central challenge in learning 1-RDMs is that a high-quality predictor must satisfy two requirements simultaneously: it should be accurate with respect to reference data, and it must produce matrices that are mathematically constrained and consistent with the electronic-structure code. The latter is crucial for practical integration.
To address this, we adopt a hybrid strategy, namely a constrained-\textsc{PhiSNet} architecture (Figure~\ref{fig:flowchart}).
First, we train the \textsc{PhiSNet} backbone, an $\mathrm{SE}(3)$-equivariant neural network,~\cite{Unke2021PhiSNet} to predict a 1-RDM $\hat{\mathbf{P}}^*$ from the molecular geometry, and then, as a second step, $\hat{\mathbf{P}}^*$ is refined by an \textit{analytical block} (Figure~\ref{fig:flowchart}A). The resulting predictor for the 1-RDM, \textsc{dm-PhiSNet},
\[
\hat{\mathbf{P}}_0=\textsc{AnalyticBlock}[\textsc{PhiSNet}(\mathbf{R})],
\]
is then used as a starting guess for the SCF calculation.
In the following subsections, we elaborate on the different components of the analytic block, data generation, and model training.

\subsection{The \textsc{AnalyticBlock}}
\label{subsec:consistency}

For restricted Kohn--Sham (RKS) solutions, physical admissibility is characterized by three coupled requirements: (i) the matrix must be Hermitian, (ii) it must conserve the electron number in the $\mathbf{S}$ metric, and (iii) it should represent a well-defined occupied subspace, reflected by generalized idempotency and a bounded occupation spectrum in an orthogonalized representation. These conditions are also the ones most directly tied to robust Roothaan--Hall fixed-point iterations. Figure~\ref{fig:flowchart}C shows how the \textsc{AnalyticBlock} imposes these conditions on the model.

\paragraph{Hermiticity.}
The 1-RDM must be Hermitian; in the real-valued setting considered here, this reduces to the symmetry,
\begin{equation}
\label{eq:herm}
\mathbf{P}=\mathbf{P}^{\mathsf T}.
\end{equation}

This is a property already fulfilled by the \textsc{PhiSNet} architecture by construction using,
\begin{equation}
\label{eq:symm-proj}
\hat{\mathbf{P}}\leftarrow\tfrac12\bigl(\hat{\mathbf{P}}+\hat{\mathbf{P}}^{\mathsf T}\bigr).
\end{equation}

\paragraph{Electron-number conservation (trace condition).}
In a non-orthogonal AO basis, the particle number is given by the $\mathbf{S}$-weighted trace,
\begin{equation}
\label{eq:trace}
\mathrm{Tr}(\mathbf{P}\mathbf{S})=N_{\mathrm e}.
\end{equation}
We impose this constraint as the first step in the \textsc{AnalyticBlock} by normalizing the \textsc{PhiSNet} density matrix (Figure~\ref{fig:flowchart}C),
\begin{equation}
\label{eq:TrConstr}
\hat{\mathbf{P}}^* \leftarrow \frac{N_{\mathrm e}}{\mathrm{Tr}(\hat{\mathbf{P}}^*\mathbf{S})}\hat{\mathbf{P}}^*.
\end{equation}

\paragraph{Generalized idempotency.}
For a closed-shell single-determinant state at self-consistency, $\mathbf{P}$ is an idempotent projector in the $\mathbf{S}$ metric,
\begin{equation}
\label{eq:idemp}
\mathbf{P}\mathbf{S}\mathbf{P}=2\mathbf{P}.
\end{equation}
This property can also be implemented as a constraint via McWeeny purification.~\cite{McWeeny1960}
We apply three iterations of the cubic purification map adapted to a non-orthogonal metric,
\begin{equation}
\label{eq:McWeeny}
\hat{\mathbf{P}}^* \leftarrow
3\,\hat{\mathbf{P}}^*\mathbf{S}\hat{\mathbf{P}}^*
-2\,\hat{\mathbf{P}}^*\mathbf{S}\hat{\mathbf{P}}^*\mathbf{S}\hat{\mathbf{P}}^*,
\end{equation}
which preserves the $\mathbf{S}$-trace and contracts residual error toward the generalized idempotency condition in Eq.~\eqref{eq:idemp}.

\paragraph{Final electron-number conservation enforcement.}
The purification process can introduce a small numerical drift in finite precision. Hence, we apply Eq.~\eqref{eq:TrConstr} once more to restore exact electron-number conservation.

The combined procedure yields an SCF-ready initializer $\hat{\mathbf{P}}_{0}=\textsc{AnalyticBlock}[\hat{\mathbf{P}}^*]$ with trace and idempotency errors at near-machine precision in all systems considered.

\subsection{Datasets}
\label{subsec:data}

All reference 1-RDMs and overlap matrices were generated using restricted Kohn--Sham DFT in PySCF~\cite{Sun} with the PBE functional and def2-type basis sets~\cite{WeigendAhlrichs2005_def2,RappoportFurche2010_propopt_basis}. All calculations use energy and density convergence thresholds substantially tighter than PySCF defaults. Geometries for ethanol were taken from the MD17 dataset.~\cite{Chmiela2017MD17} For H$_2$O, NH$_3$, CH$_4$, and HF, structures were generated using ASE~\cite{Hjorth2017} and \textit{ab initio} molecular-dynamics simulations using the FHI-aims package.~\cite{FHIaims2009}
For NO$_3^-$, structures were obtained using the GFN2-xTB semi-empirical method~\cite{Bannwarth2019GFN2xTB} followed by DFT refinement.
Further details on sampling, dataset sizes, matrix dimensions, train/validation/test splits, and SCF thresholds are provided in the Supporting Information.

\subsection{Model and training}
\label{subsec:training}

\paragraph{Backbone architecture.}
The \textsc{dm-PhiSNet} model uses the \textsc{PhiSNet}~\cite{Unke2021PhiSNet} architecture, an $\mathrm{SE}(3)$-equivariant message-passing network whose internal features transform as irreducible representations (irreps). Tensor-product mixing and equivariant nonlinearities yield rotationally consistent predictions of AO-block matrices from molecular configurations. We adapt the output heads from Hamiltonian prediction $\hat{\mathbf{H}}$ to direct prediction of the AO 1-RDM, $\hat{\mathbf{P}}$.

\begin{figure}
\centering
\includegraphics[width=\columnwidth]{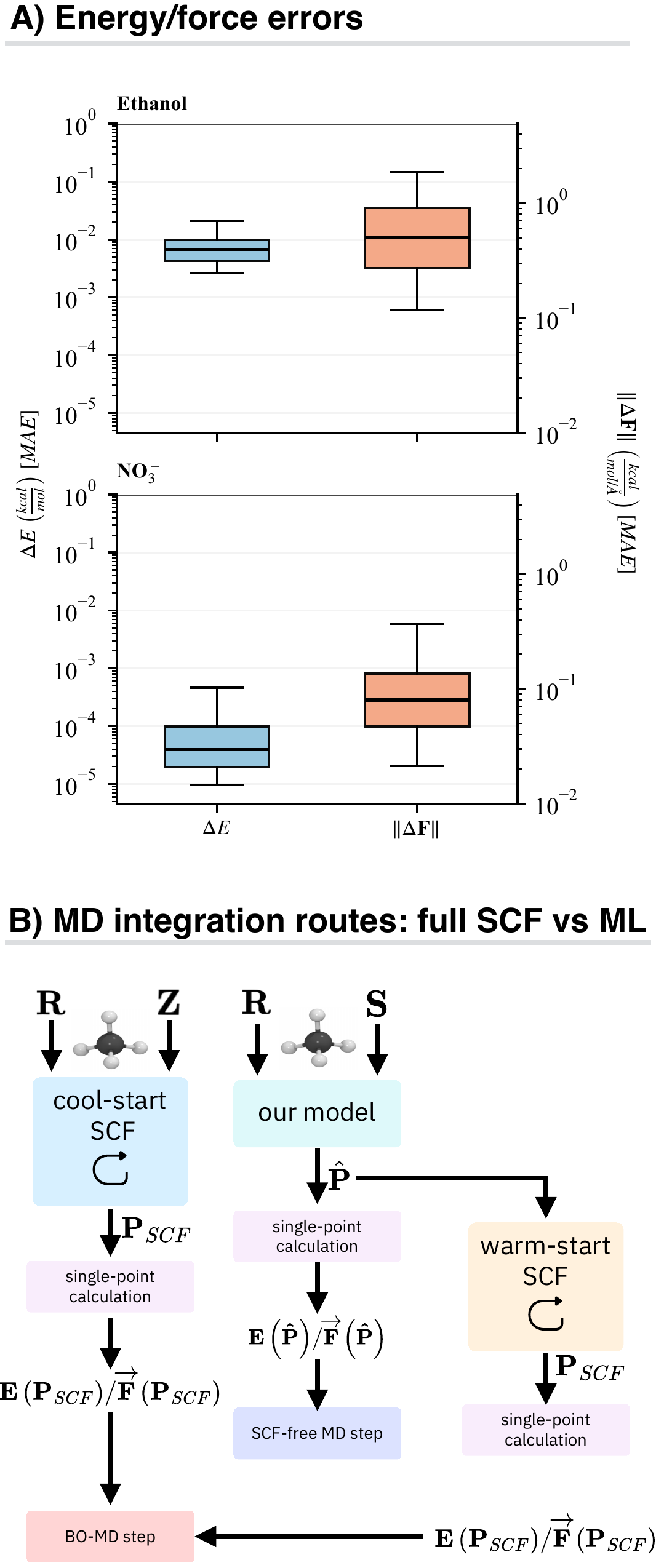}
\caption{\textbf{(A)} Distributions of single-point errors in total energy $\Delta E$ (left axis) and force norms $\|\Delta\mathbf{F}\|$ (right axis) computed directly from the predicted 1-RDM $\hat{\mathbf{P}}_0$, for ethanol (top panel) and $\mathrm{NO_3^-}$ (bottom panel).
\textbf{(B)} Conceptual integration into MD: conventional SCF versus using $\hat{\mathbf{P}}_0$ either (i) directly for an SCF-free MD step (as a CP-style electronic initializer) or (ii) as a warm start for accelerated SCF refinement for BO-MD.}
\label{fig:energy_force}
\end{figure}

\paragraph{Two-stage training protocol.}
Training proceeds in two phases. In Stage~1, we minimize a global mean-squared error between predicted and reference 1-RDMs to learn a stable geometry-to-1-RDM map.
In Stage~2, we introduce auxiliary loss terms associated with the elements in the \textsc{AnalyticBlock} (trace and generalized idempotency), together with occupation-spectrum violations. Constraint terms contribute to the gradient only when their diagnostic value exceeds $10^{-6}$, preventing over-penalization once satisfied.
Because AO matrices have a natural block structure by subshell ordering, we additionally use a block-wise objective that remains active throughout Stage~2 to reduce localized errors in high-angular-momentum blocks. Precise definitions of the loss components, schedules, and training hyperparameters are provided in the Supporting Information.

\section{Results}
\label{Results}

In this section, we present a general overview of the SCF, energy, and force benchmarks of our model. A more detailed numerical report of the model's performance is available in the Supporting Information.

\subsection{Global errors and performance as SCF initializer}
\label{subsec:scferformance}

Modern quantum-chemistry packages provide built-in initial guesses. The most common include the minimal atomic orbital (MINAO) guess,~\cite{Almlof1982,VanLenthe2006} the \texttt{atom} guess (also known as the superposition of atomic densities, SAD),~\cite{VanLenthe2006} and the H\"uckel initializer.~\cite{huckel1931benzolI,huckel1931benzolII} These serve as the main baselines in our SCF performance analysis.

The main metric used to estimate the performance of an initial DM guess is the number of steps required by the SCF algorithm to converge the electronic-structure calculation. Figure~\ref{fig:scf_reductions} summarizes the average SCF iteration counts for all benchmarked systems by comparing the performance of our predicted 1-RDMs $\hat{\mathbf{P}}_0$ against the conventional initial guesses mentioned above.
Our predictions for all six molecules consistently lead to a substantial reduction of more than 49\% in the number of SCF cycles.

A particular emphasis is placed on hydrogen fluoride (HF). Although HF is a simple diatomic molecule, it is strongly polar due to the extreme electronegativity of fluorine, which concentrates electronic density around its center.
The charge asymmetry in this system highlights the importance of a high-quality initial guess. When the initial guess is poor, it can lead to inefficient or unstable SCF convergence. In contrast, our model achieves the most substantial acceleration for all the tested out-of-equilibrium configurations, reducing iteration counts by over 80\%.

All results reported in Figure~\ref{fig:scf_reductions} correspond to the direct inversion of the iterative subspace (DIIS),~\cite{Pulay1980DIIS,Pulay1982DIIS} which is arguably the most widely used SCF solver in modern DFT applications. All SCF benchmarks were carried out under deliberately tight numerical thresholds (energy: $10^{-9}$~Ha, density: $10^{-8}$, gradient: $10^{-6}$~Ha/Bohr). These criteria are more stringent than the typical default settings in many electronic-structure workflows. Therefore, they yield a more discriminating test of SCF initial guesses. In our experience, looser tolerances can reduce iteration counts across methods by allowing termination before the fixed point is reached with high precision. By enforcing stricter convergence, we ensure that differences in SCF effort reflect genuine improvements in proximity to the converged solution rather than artifacts of permissive stopping criteria. All initial guesses are evaluated under identical settings.

The mean absolute errors (MAEs) of the predicted 1-RDMs across all benchmark systems are summarized in Table~\ref{tab:dm_mae}. Notably, the MAE varies substantially across systems, with ethanol showing a larger entrywise error than the smaller molecules. This spread is expected from the fact that ethanol's DM is substantially larger than that of ten-electron systems. Therefore, it poses a higher-complexity matrix to learn (larger $N_{\mathrm{AO}}$ and more heterogeneous subshell blocks). On the other hand, the MAE by itself is not an informative metric of downstream SCF performance, nor of energy and force accuracy.

In practice, these are often governed by whether the predicted 1-RDM is properly constrained in the AO metric (electron number, generalized idempotency, and a sensible occupation spectrum). This is evident in Figure~\ref{fig:scf_reductions}, where despite the higher MAE, the model still outperforms the conventional initial guesses by $\approx 50\%$ in the number of SCF cycles.

A report on reductions in SCF iterations and wall-clock time is shown in the Supporting Information.

\begin{table}[t]
\centering
\caption{MAEs of the predicted 1-RDMs relative to SCF reference values, averaged over 500 test instances.}
\label{tab:dm_mae}
\renewcommand{\arraystretch}{1.25}
\begin{tabular}{l c}
\toprule
\textbf{Molecule} &
\multicolumn{1}{c}{\textbf{$\lVert \Delta P\rVert_{\mathrm{MAE}}$ [$\times 10^{-6}$]}} \\
\midrule
H$_2$O       & 3.124  \\
CH$_4$       & 2.181  \\
NH$_3$       & 2.407  \\
HF           & 0.357  \\
C$_2$H$_5$OH & 53.390 \\
NO$_3^{-}$   & 3.692  \\
\bottomrule
\end{tabular}
\end{table}

\subsection{Energy and force predictions from $\hat{\mathbf{P}}_0$}
\label{subsec:energy_force}

A commonly used rule of thumb for \emph{chemical accuracy} is an error of $\sim 1\,\mathrm{kcal/mol}$ in \emph{relative} electronic energies (e.g., reaction energies and barrier heights),~\cite{Loos2018_theoretical_00} corresponding to $4.3\times 10^{-2}\,\mathrm{eV}$.
To assess whether our predicted 1-RDMs capture an electronically meaningful structure that goes beyond their role as SCF initial guesses, we compute total energies and Hellmann--Feynman forces~\cite{Hellmann1937_einfuehrung,Feynman1939_forces_molecules} \emph{directly} from the predicted 1-RDM ($\hat{\mathbf{P}}_0$), without performing any SCF iterations. These results were compared against values obtained from DFT-converged 1-RDMs.
As an additional solver-relevance diagnostic, the predicted 1-RDMs yield uniformly small orbital-energy errors across all six systems (per-orbital MAEs $\ll 1$~kcal/mol over 500 test configurations; see Sec.~S5 in the Supporting Information).

Figure~\ref{fig:energy_force} shows the total-energy and force errors for ethanol and the nitrate anion. While ethanol is the most flexible and chemically diverse system in our benchmark set---and therefore the most challenging case---it still reaches a total-energy MAE well below the chemical-accuracy threshold.
Moreover, when used as an SCF initial guess, it reduces the number of convergence cycles by about 49.7\% (Figure~\ref{fig:scf_reductions}). Taken together, these results indicate that the predicted 1-RDM $\hat{\mathbf{P}}_0$ provides a reliable warm start for ethanol that can be brought to full self-consistency with only a few additional SCF iterations.
Specifically, the one-shot energies (e.g., $E[\hat{\mathbf{P}}_0]$) for ethanol and $\mathrm{NO_3^-}$ are $\approx 6\times10^{-3}$~kcal~mol$^{-1}$ and $\approx 2\times10^{-5}$~kcal~mol$^{-1}$, respectively. The remaining ten-electron systems report total-energy MAEs that lie even further below these values.

An additional promising result is the accuracy of one-shot atomic forces for all systems in our benchmark, in the range of 0.02--0.2~kcal~mol$^{-1}$~\AA$^{-1}$.
This accuracy is competitive with modern machine-learned force fields trained explicitly on force labels.~\cite{unke2021spookynet,Batzner2022NequIP,Batatia2022MACE,2025_DeltaML_Cazares,Unke.MLFF.ChemRev2020}
Importantly, these force errors are achieved \emph{without any force supervision during training}: the forces emerge solely from reconstructing a physically consistent electronic structure via the predicted 1-RDM.
This indicates that enforcing the mathematical admissibility of the density matrix is sufficient to reconstruct consistent atomic forces.
As a consequence, the predicted electronic states for the smaller systems could be suitable for downstream quantum-mechanical workflows such as molecular dynamics (MD). In particular, they could serve as initializers for Car--Parrinello (CP)-type~\cite{CarParrinello1985} MD, where reducing the initial electronic equilibration overhead is beneficial. Figure~\ref{fig:energy_force}B illustrates the potential procedures suggested for integrating our predicted 1-RDMs into MD workflows.

\subsection{Discussion}
\label{subsec:discussion}

The N-representability problem has been a notorious topic in electronic-structure theory.~\cite{NRC1995MathChallengesChem} For the one-electron case, physical consistency is equivalent to N-representability. The challenge of directly predicting physically consistent 1-RDMs has been bypassed using relatively involved strategies in previous ML works.~\cite{Rana2024Learning1RDM,Hazra2024Predicting1RDM}
The downstream analytic block described here offers a lightweight (Figure~\ref{fig:flowchart}C) and versatile alternative for straightforwardly predicting physically consistent 1-RDMs. This translates explicitly into significant improvements in electronic-structure correctness, thus delivering accurate one-shot energies and forces and consistently reducing the required SCF cycles to reach electronic-structure convergence.

Particularly for ethanol, the predicted molecular energy MAE does not exceed $10^{-2}$~kcal/mol (Figure~\ref{fig:energy_force}), almost one order of magnitude below the threshold chosen by Liu \textit{et al.}~\cite{Liu2025MachineLearnedFockMatrix} to learn Fock matrices for considerably smaller molecules. Similar results are obtained when explicitly comparing occupied orbital-energy MAEs against Liu \textit{et al.}'s work on the H$_2$O molecule.
As mentioned in the previous section (\ref{subsec:energy_force}), our reported force accuracy is another salient consequence of enforced physical consistency alone, without explicit force-supervised training. Hazra \textit{et al.}~\cite{Hazra2024Predicting1RDM} report $x$- and $y$-component force MAEs for S$_2$O of $62~\mathrm{meV}/\AA$ and $22~\mathrm{meV}/\AA$, respectively. These results were improved using a different trace-electron-count correction than the one described in Sec.~\ref{sec:methods}. In contrast, for ethanol, our most complex model, we achieve (Figure~\ref{fig:energy_force}) a norm force MAE of $\approx 1.6~\mathrm{meV}/\AA$, representing a substantial improvement over their work.
Naturally, for systems simpler than ethanol, our energy and force predictions are even more accurate. More importantly, our model is naturally applicable to predicting physically consistent 1-RDMs for larger closed-shell systems by providing spatial and AO-basis information.

\section{Conclusion}
\label{Conclusion}

In this work, we devise \textsc{dm-PhiSNet} to predict physically consistent AO 1-RDMs by combining the \textsc{PhiSNet} backbone with a lightweight analytic refinement block at negligible overhead ($\lesssim 0.14\%$ of a network evaluation).
A central outcome is that \emph{physically constrained} 1-RDMs---in particular, electron-number conservation, generalized idempotency in the AO metric, and a well-behaved occupation spectrum---govern SCF robustness more strongly than minimizing raw entrywise prediction error.
Across six chemically distinct systems, \textsc{dm-PhiSNet} reduces SCF iteration counts by 49--81\% relative to standard initial guesses, while simultaneously yielding reliable one-shot atomic forces even though no force information is used during training.

These findings demonstrate that enforcing the \emph{right physical constraints} is sufficient to transform equivariant neural predictions into solver-ready electronic states. By projecting learned 1-RDMs onto the admissible manifold at essentially zero cost, \textsc{dm-PhiSNet} provides a principled and practical bridge between machine learning and \textit{ab initio} electronic-structure codes, turning machine learning from a standalone predictor into a scalable accelerator for routine quantum-chemical workflows in larger, more complex closed-shell systems.

\section*{Acknowledgements}
\label{acknowledgements}

H.E.S. acknowledges support from CONAHCYT/SECIHTI-Mexico under Project CF-2023-I-468, DGTIC-UNAM under Project LANCAD-UNAM-DGTIC-419, and DGAPA-UNAM PAPIIT Nos.\ IA106023 and IA105625. A.A.G. gratefully thanks CONAHCYT/SECIHTI-Mexico for PhD scholarship No.\ 957574. Z.Y.Y.R. expresses gratitude to CONAHCYT/SECIHTI-Mexico for the financial support provided during his doctoral studies. We acknowledge Carlos Ernesto L\'opez Natar\'en for helping with the high-performance computing infrastructure.

\section*{Conflict of Interest}
\label{conflict}

The authors declare no conflicts of interest.


\bibliography{references.bib}
%
%
%
%


\end{document}